\documentclass{article}
\usepackage{amsmath,amssymb,graphicx,color,graphics,verbatim}
\usepackage[margin=1.25in]{geometry}
\usepackage{hyperref}
\usepackage[utf8]{inputenc}

\newcommand{\nc}{\newcommand}

\nc{\beq}{\begin{equation}}
\nc{\eeq}{\end{equation}}
\nc{\barray}{\begin{eqnarray}}
\nc{\earray}{\end{eqnarray}}
\nc{\barrayn}{\begin{eqnarray*}}
\nc{\earrayn}{\end{eqnarray*}}
\nc{\bcenter}{\begin{center}}
\nc{\ecenter}{\end{center}}
\nc{\ket}[1]{| #1 \rangle}
\nc{\bra}[1]{\langle #1 |}
\nc{\0}{\ket{0}}
\nc{\mc}{\mathcal}
\nc{\er}[1]{(\ref{eq:#1})}
\nc{\onehalf}{\frac{1}{2}}
\nc{\partialbar}{\bar{\partial}}
\nc{\psit}{\widetilde{\psi}}
\nc{\Tr}{\mbox{Tr}}
\nc{\hc}{\mbox{H.c.}}
\nc{\ev}{\;\mathrm{eV}}
\nc{\mev}{\;\mathrm{MeV}}
\nc{\gev}{\;\mathrm{GeV}}
\nc{\tev}{\;\mathrm{TeV}}

\def\chii0{\chi_i^0}
\def\chij0{\chi_j^0}

\newcommand{\gsim}{\lower.7ex\hbox{$\;\stackrel{\textstyle>}{\sim}\;$}}
\newcommand{\lsim}{\lower.7ex\hbox{$\;\stackrel{\textstyle<}{\sim}\;$}}
\nc{\ttbar}{t\bar t}

\newcommand{\fref}[1]{Fig.~\ref{f.#1}}
\newcommand\snowmass{\begin{center}\rule[-0.2in]{\hsize}{0.01in}\\\rule{\hsize}{0.01in}\\
\vskip 0.1in Submitted to the  Proceedings of the US Community Study\\ 
on the Future of Particle Physics (Snowmass 2021)\\ 
\rule{\hsize}{0.01in}\\\rule[+0.2in]{\hsize}{0.01in} \end{center}}

\def\Title#1{\begin{center} {\LARGE #1 } \end{center}}
\def\Author#1{\begin{center}{ \sc #1} \end{center}}
\def\Address#1{\begin{center}{ \it #1} \end{center}}

\newenvironment{Abstract}{\begin{quotation} \begin{center}
                       ABSTRACT
     \end{center}\bigskip  }{\end{quotation}}

\begin{document}

\Title{Exotic Higgs Decays to Four Taus at Future Electron-Positron Colliders}

\bigskip 

\Author{Jessie Shelton and Dong Xu}

\medskip

\Address{Illinois Center for the Advanced Studies of the Universe, Department of Physics, \\University of Illinois at Urbana-Champaign}

\medskip

 \begin{Abstract}
\noindent  We study the prospects for the exotic Higgs decay mode $h\to ss\to 4\tau$ at planned electron-positron colliders in the case where the beyond-the-Standard Model particle $s$ is too light to decay into $b \bar b$ pairs.  We find that with 5 ab$^{-1}$ of  unpolarized collisions at 240 GeV, the branching ratio into this final state can be constrained to be below $\sim 1.3\times 10^{-4}$  at 95\% CL, depending on the mass of $s$ and the assumed tracking efficiency.   
\end{Abstract}

\snowmass

\section{Introduction}

The discovery of the Higgs boson at the Large Hadron Collider (LHC) has provided major opportunities to test its properties and shed light on possible new physics that could be produced in its decays.  
The vast samples of Higgs bosons that will be produced at the upcoming high luminosity run of the LHC can allow experimental probes of tiny exotic Higgs branching fractions, provided that the resulting signature is sufficiently clean to be separated from the hadronic backgrounds.  The exemplar of such a decay is provided by $h\to Z_D Z_D \to 4 \ell$, for which HL-LHC sensitivity is estimated to reach $Br (h\to Z_D Z_D) < 10^{-6}$ \cite{Curtin:2014cca}. 

However, many final states of particularly high interest have higher backgrounds and thus present more challenging targets at the LHC.
 Planned electron-positron colliders, operating as Higgs factories, will provide substantial complementary information about the Higgs boson and its properties by providing a large statistical sample of low-background events.  The benefit of such clean events for precision studies of Higgs couplings to SM particles has long been appreciated; recent work is summarized with up-to-date forecasts in \cite{deBlas:2019rxi}.  Such environments will also allow sensitive probes of exotic Higgs branching ratios into final states that are challenging to study at the HL-LHC, in particular all-hadronic and/or high multiplicity final states \cite{Liu:2016zki}.

One particularly interesting class of signatures is provided by Higgs decays to pairs of (pseudo-)scalars that decay via Higgs portal interactions.  Such exotic decays featuring (pseudo-)scalars are motivated by a wide range of theories, from relaxions \cite{Flacke:2016szy,Frugiuele:2018coc} to models of neutral naturalness (in which case the scalar is often long-lived) \cite{Craig:2015pha,Curtin:2015fna} to models of thermal DM \cite{Ipek:2014gua,Martin:2014sxa,Evans:2017kti}.  A light, visibly decaying Higgs portal scalar can still drive the electroweak phase transition strongly first-order in narrow but still viable portions of parameter space \cite{Kozaczuk:2019pet,Carena:2019une}.
Scalars that decay via Higgs portal interactions dominantly decay to final states that can be challenging at high-background hadron colliders. For such models, the somewhat smaller but much lower-background Higgs samples provided by planned electron-positron colliders, operating as Higgs factories, will be particularly valuable for discovery or exclusion. When the scalar has $m_s > 2 m_b$, the dominant exotic branching fraction is $h\to ss\to 4b$,
for which Ref.~\cite{Liu:2016zki} estimates that future electron-positron colliders will constrain $Br (h\to 4b) < \mathrm{few} \times 10^{-4}$ at the 95\% confidence level, for $15\gev \lesssim m_s\lesssim 60$ GeV.

For $m_s<10$ GeV, decays into $b\bar b$ are kinematically forbidden and the dominant branching fraction is instead $h\to 4\tau$, where the hierarchy $m_s \ll m_h$ will collimate the $s\to\tau\tau$ decay. The $\tau$ dominantly decays hadronically, making these signatures challenging at hadron colliders, and at the LHC, the leading constraint on decays to Higgs-mixed scalars in this mass range comes from $h\to 4\mu$ \cite{CMS:2020bni,ATLAS:2021ske}.  There is no current survey of sensitivities to Higgs exotic decay modes for $m_s < 10$ GeV at lepton colliders.  Previous studies of $h\to 4\tau$ at lepton colliders include the searches carried out at LEP \cite{Abbiendi:2002in,Schael:2010aw}, and an ILC study performed in the context of the NMSSM \cite{Liu:2013gea}.  The aim of the current work is to provide an up-to-date estimate of planned electron-positron collider reach for the decay $h\to ss \to 4\tau$, which will substantially clarify the discovery prospects for challenging Higgs decay modes below the $b\bar b$ threshhold.

\section{Colliders, Signal, and Background}

We consider a reference running scenario with an integrated luminosity of $\mathcal{L} = 5\,\mathrm{ab}^{-1}$ collected in unpolarized electron-positron collisions at a 240 GeV CM energy, which, based on the proposals summarized in \cite{deBlas:2019rxi}, is a useful common reference  for both CEPC \cite{CEPCStudyGroup:2018ghi} and FCC-ee \cite{FCC:2018byv}.  This reference scenario is also broadly representative of ILC's planned initial run, 2 ab$^{-1}$  at 250 GeV with polarized beams, which would produce a little more than half as many Higgs events
\cite{Bambade:2019fyw,LCCPhysicsWorkingGroup:2019fvj}.

Our signal process is $e^+e^-\to Zh$, followed by $Z\to \ell\ell$ and $h\to ss\to 4\tau$.  We are primarily interested in the mass range 5 GeV$< m_s<10$ GeV, where the two $s\to 2\tau$ decays  are collimated in the detector.
The dominant background process is hadronic SM Higgs decays, $e^+e^-\to Z h$, $h\to jj$; we additionally consider the non-Higgs processes (i) $e^+e^-\to Z jj$ and (ii) $e^+e^-\to e^+ e^-Z$ followed by $Z\to jj$.  

Both signal and SM background event samples are generated using Madgraph 5 \cite{Alwall:2014hca} interfaced to Pythia 8 \cite{Sjostrand:2007gs}.   Signal $h\to ss\to 4\tau$ events are generated using the \texttt{HAHM}$\_$\texttt{variableMW}$\_$\texttt{v2} Madgraph user model \cite{Curtin:2014cca}.  
From Madgraph, the cross section for $e^+e^-\to Zh$, followed by $Z\to \ell\ell$, is $\sigma_{hZl}=8.0$ fb (including $Z$ decays to both electrons and muons).
The signal cross section is then $\sigma_{hZl} \times Br(h\to ss\to 4\tau)$, while the background cross-section is $\sigma_{hZl} \times Br(h\to jj)$.  
We add $h\to bb$, $cc$, and $gg$ branching fractions to obtain $Br(h\to jj) = 0.693$~\cite{yr4}.  

We also use the cross-sections computed by Madgraph for the two non-Higgs background processes mentioned above.  We find these backgrounds are efficiently vetoed by simple $Z$ and recoil mass cuts, described below.

\section{Event Analysis}

We require final state particles to have $|\cos\theta| < 0.99$ and $|\vec p| > 0.25$ GeV in order to be identified by the detector. 
In order to isolate $Zh$ events, we require the presence of two oppositely charged electrons or muons that satisfy $|m_{\ell\ell}-m_Z|<10$ GeV.  The recoil mass in the event,
\beq
m_{rec} \equiv \sqrt{s - 2 \sqrt{s} E_{\ell\ell} + m_{\ell\ell}^2},
\eeq
is required to satisfy $|m_{rec} - m_h| < 5$ GeV.  These two cuts effectively eliminate the non-Higgs backgrounds.

After identifying the leptons that reconstruct the $Z$ boson, we cluster the remaining final state particles into jets using the \texttt{ee genkt} algorithm with a large radius parameter $R = 1.3$ in  FastJet  \cite{Cacciari:2011ma}.  We require at least two jets with $|\vec p|> 5 $ GeV.  

\begin{figure}[t]
\begin{center}
\begin{tabular}{cc}
\includegraphics[width=7.6cm]{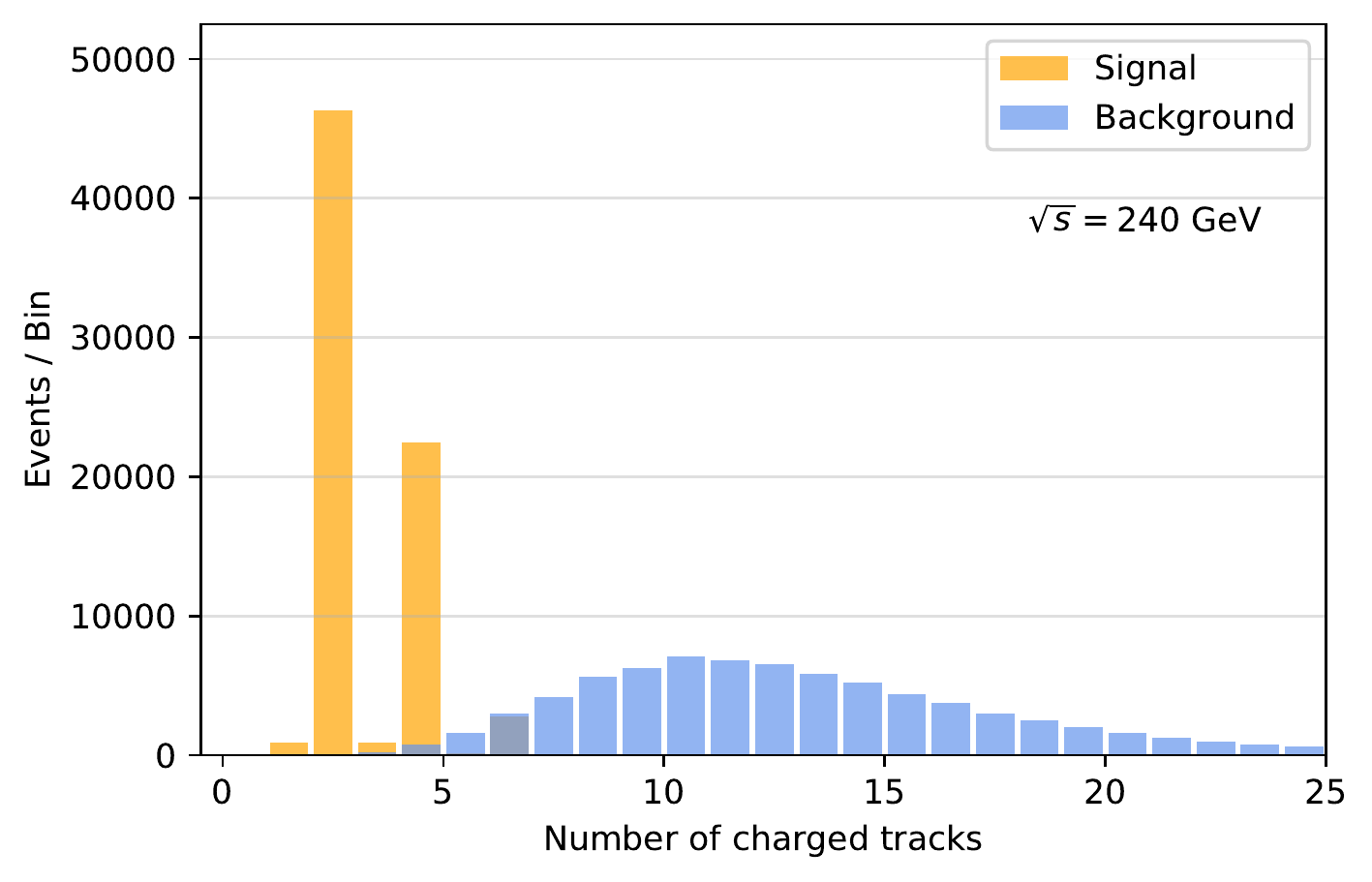}
&
\includegraphics[width=7.7cm]{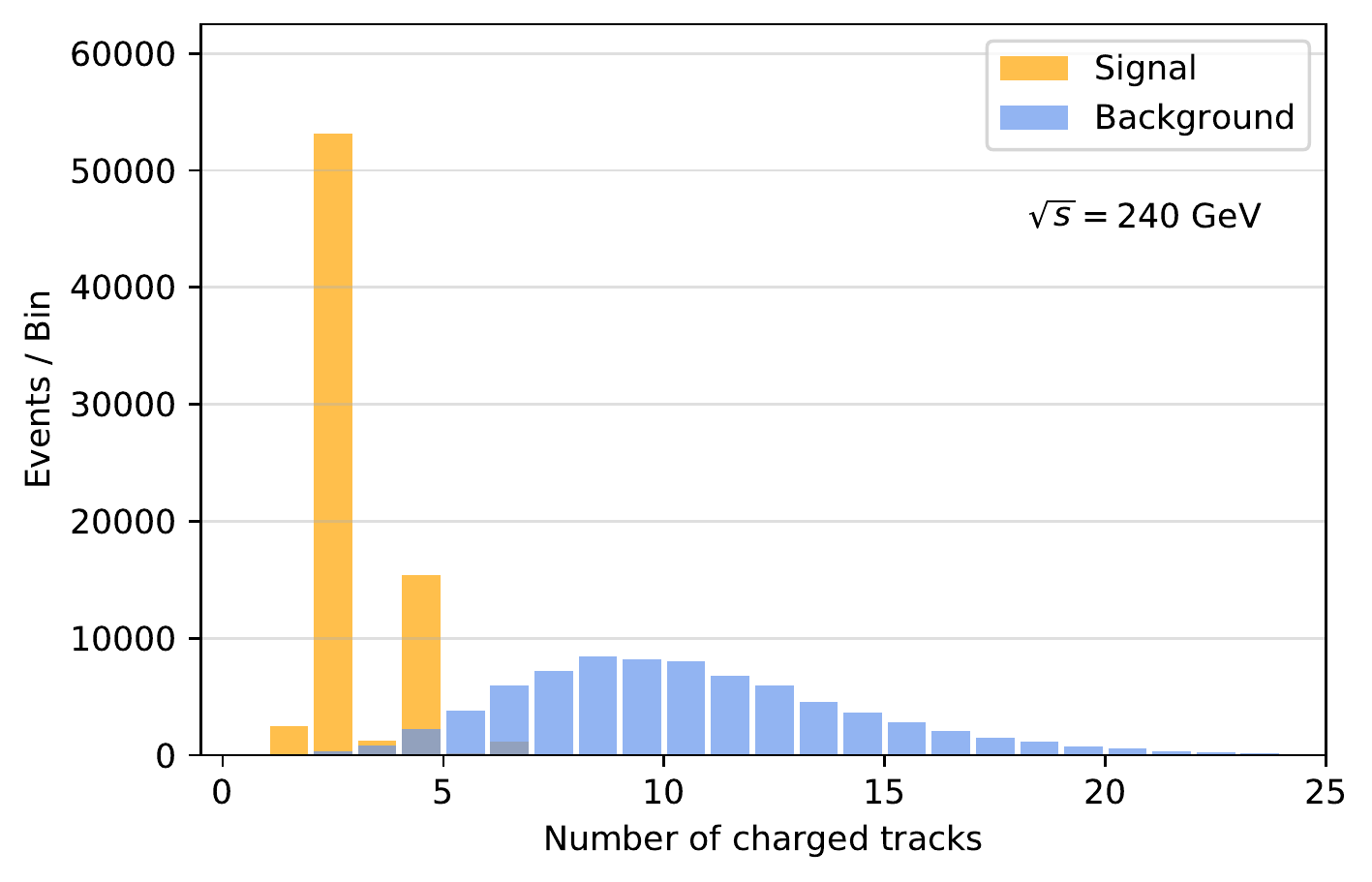}
\end{tabular}
\end{center}
\caption{{\small Number of charged tracks in the hardest  (left) and second hardest  (right) jets, for SM $h\to jj$ (blue)  and BSM $h\to 4\tau$ (orange), with $m_s =7.5$ GeV.  Events shown are required to pass $Z$ reconstruction and recoil mass cuts and have at least two jets. Samples of $10^5$ events are simulated for signal and background separately.}}
\label{f.CTN}
\end{figure}

The single most useful observable for separating signal and background is the charged track multiplicity within the two hardest fat jets, shown in Fig.~\ref{f.CTN}.  
Requiring that each of the two most energetic jets have exactly two charged tracks is our primary analysis cut. We additionally require that, if there is a third jet in the event, it must have no charged tracks.  Background processes with Higgs decays to QCD-charged particles may have hadronic final state radiation, while for signal, these ``jets'' are often photons; thus in a full analysis this track veto might in a full analysis be replaced by a photon identification criterion.  These two track-counting cuts are, by themselves, sufficient to almost entirely suppress the background while retaining more than a third of the signal events, as we show in Table~\ref{t.signal_bkg_percentage} for the mass point $m_s = 7.5$ GeV.   After these cuts, the SM background has been suppressed to such a level that additional cuts on e.g. the invariant mass of the fat jets will reduce sensitivity by chipping away at the signal efficiency.
%
For this mass point, we obtain the 95\% C.L. upper limit $Br(h\to 4\tau) <1.27\times10^{-4}$ using the CLs method \cite{read2002presentation}. Upper limits for other values of $m_s$ are shown in \fref{limits_points}.
%
\begin{table}[htbp]
\begin{center}
\begin{tabular}{| l ||  c | c | c | c ||}
\hline 
& Simulated events & Selected events & Efficiency & Events in 5 ab$^{-1}$ \\
\hline 
Signal ($m_s = 7.5$ GeV) & $5\times10^5$ & 168995 & $0.338 $ & 27039.2 $\times Br$ \\
\hline
SM background & $100\times10^5$ & 54 & $5.4 \times 10^{-6}$ & 0.2997 \\
\hline
\hline
Signal, $\epsilon_{tk}=0.99$  & $5\times10^5$   & 162558 & $0.3251$ & 26009.3 $\times Br$ \\
\hline
SM background, $\epsilon_{tk}=0.99$ &  $100\times10^5$ & 63 & $6.3 \times 10^{-6}$ & 0.3497 \\
\hline
\end{tabular}
\end{center}
\caption{{\small Summary of signal and background efficiencies for $m_s = 7.5$ GeV, after applying the cuts described in the text. The last column shows the expected number of events at future $e^+ e^-$ colliders with $\sqrt{s}=240$ GeV and $\mathcal{L}=5\;\mathrm{ab}^{-1}$. The symbol ``Br" represents the Higgs exotic decay branching fraction $Br (h\to4\tau)$.  The top two rows show results with perfect track reconstruction, while the bottom two rows show results for the same signal mass point assuming a flat track reconstruction efficiency $\epsilon_{tk} =0.99$.}}
\label{t.signal_bkg_percentage}
\end{table}

So far we have assumed perfect tracking efficiency for prompt charged particles satisfying the basic momentum and angular acceptance criteria.  While we expect  tracking efficiency will be excellent at future detectors, of course it will not be perfect.   However, as Fig.~\ref{f.CTN} suggests, the SM and signal track multiplicity distributions are sufficiently distinct that finite tracking efficiency will not substantially alter our sensitivity estimate.  We show in Table~\ref{t.signal_bkg_percentage} and \fref{limits_points} how our results change with an assumed $99\%$ efficiency for track identification.  The finite track efficiency weakens our estimated sensitivity at the $\sim 5\%$ level.

With finite tracking efficiency, the SM decay $h\to \tau\tau$ can also contribute to the signal final state when two three-prong tau decays are each mis-reconstructed as two-track final states. Assuming a $99\%$ efficiency for track identification, the effective branching ratio into the signal final state from $h\to \tau\tau$ with one missed track on each side is $Br(h\to 2\tau)_{misID}\equiv Br(h\to 2\tau) \times Br (\tau\to 3h^\pm)^2\times .03^2$, where $Br (h\to 2\tau) = 0.06272$ \cite{yr4}, the three-prong $\tau$ branching ratio $ Br (\tau\to 3h^\pm) = .1519$ \cite{ParticleDataGroup:2020ssz}, and the product is $Br (h\to2\tau)_{misID} = 1.3\times 10^{-6}$.  This mis-ID rate is two orders of magnitude smaller than the branching ratios for $h\to 4\tau$ to which our analysis finds sensitivity, so we conclude that as long as track efficiency remains high, the $h\to \tau\tau$ background will not be a substantial contribution to the signal final state.

\begin{figure}[t]
\begin{center}
\includegraphics[width=10.0cm]{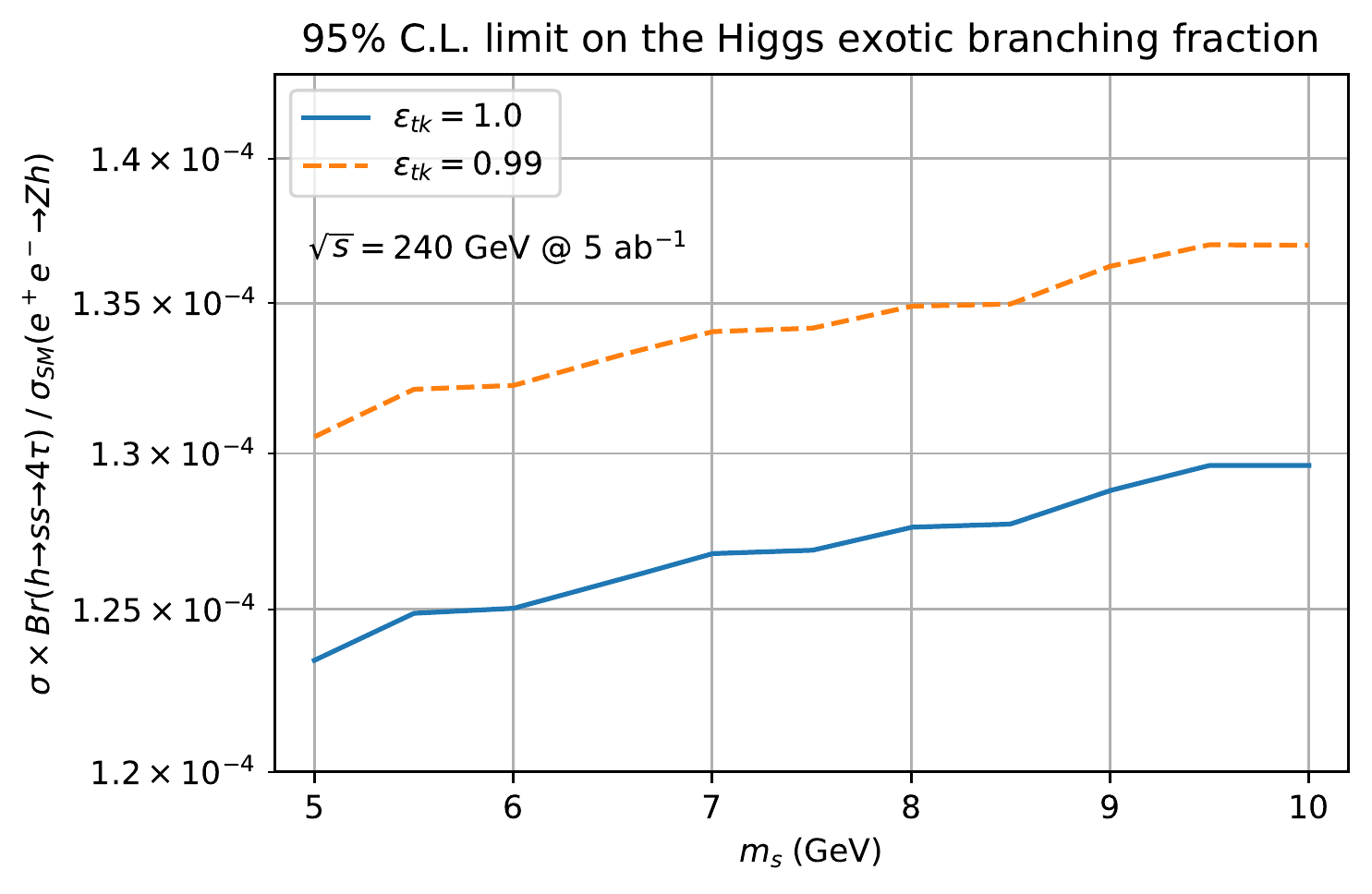}
\end{center}
\caption{{\small Estimated 95$\%$ C.L. upper limits on  $Br(h\to 4\tau)$ at future $e^+ e^-$ colliders with $\sqrt{s}=240$ GeV and $\mathcal{L}=5\;\mathrm{ab}^{-1}$, assuming perfect tracking efficiency (blue, solid) and 99\% tracking efficiency (orange, dashed) for particles with $|\vec p| > 0.25$ GeV and $|\cos\theta|<0.99$.}}
\label{f.limits_points}
\end{figure}

\section{Summary}

We have estimated the sensitivity of future lepton colliders to the exotic Higgs decay $h\to ss\to 4\tau$ in the regime where the BSM (pseudo-)scalar $s$ is light, $5\gev < m_s< 10 \gev$.  This decay mode is the dominant branching fraction for Yukawa-coupled scalars below the $b\bar b$ threshold, and is a state of high interest for models featuring light spin-zero particles that couple to the SM through mixing with the Higgs boson. We find that with 5 ab$^{-1}$ of data at 240 GeV, branching fractions into this final state can be tested at the level of $\sim 10^{-4}$; our full results are shown in \fref{limits_points}.  This $\mathcal{O} (10^{-4})$ sensitivity is broadly similar to the sensitivity found for a range of final states at higher values of $m_s$ in \cite{Liu:2016zki}.  Our analysis relies primarily on simple track counting cuts, and thus will depend on the tracking efficiency that detectors will be able to achieve, but we expect that the tracking cuts will remain powerful tools to separate signal and background in realistic environments.

\bibliography{h4tau}
\bibliographystyle{JHEP}


\end{document}